\documentclass[pra,twocolumn,superscriptaddress,showpacs,preprintnumbers,eqsecnum,amsmath,amssymb]{revtex4}

\begin{document}
\title{Encryption via Entangled states belonging to Mutually Unbiased Bases}
\author{M. Revzen}
\affiliation {Department of Physics, Technion - Israel Institute of
Technology, Haifa 32000, Israel}
\author{ F. C. Khanna}
\affiliation {Department of physics, University of Alberta, Edmonton,
Alberta, Canada T6G 2J1}

\date{\today}

\begin{abstract}
We consider particular entanglement of two particles whose state vectors are in bases that are mutually unbiased (MUB), i.e. "that exhibit maximum degree of incompatibility" (J.Schwinger,Nat. Ac. Sci. (USA), 1960) ). We use this link between  entanglement and MUB  to outline a protocol for secure key distribution among the parties that share these entangled states. The analysis leads to an association of entangled states and states in an MUB set: both carry the same labels.
\end{abstract}
\pacs{03.65-w,03.67Dd,89.70.+c}

\maketitle


\section{Introduction}

Two orthonormal vector bases, $ {\cal B}_{1},\;{\cal B}_{2}$, are said to be mutually unbiased (MUB) iff
\begin{equation}
\forall\; |u_1\rangle,\;|u_2\rangle\;\epsilon\; {\cal B}_{1},\;{\cal B}_{2}\;\;resp.\;\; |\langle u_1|u_2 \rangle |=constant,
\end{equation}
 i.e. the absolute value of the scalar product of vectors from different bases is independent of the vectorial labels within either basis. This implies that if the state vector is measured to be in one of the states e.g. $|u_1\rangle$ of the base
${\cal B}_{1}$,
it is equally likely to be in any of the states $|u_2\rangle$ of the base ${\cal B}_2$. The first to emphasize that there are  more than two such bases "that exhibit maximum degree of incompatibility"  i.e. more than just the pair of conjugate bases such as $|x \rangle,$  (spatial coordinates) and $ |k \rangle$ (momentum representation basis) was Schwinger \cite{schwinger}. The infromation theoretic oriented term "mutually unbiased bases" (MUB) is due to Wootters \cite{wootters1}. Wootters and coworkers \cite{wootters1,wootters2,wootters3} related the MUB's to lines in phase space: the vector/state that satisfy the equation,
\begin{equation} \label{one}
 (\hat {p}-b\hat {x} )|1;b,c\rangle\;=\;c | 1;b,c\rangle,\;\;-\infty< c,b<\infty,
\end{equation}
This state may be viewed as lines in phase space in as much as the Wigner function of such state is given by (cf. Eq.(\ref{mub1}),  Appendix A),
\begin{equation}\label{wigner1}
W_{|1;b,c \rangle}(q,p)\;=\;\delta(p-bq-c).
\end{equation}

\noindent The index 1 among the state's labels, Eq.(\ref{one}), denotes a one particle state (correspondingly 2 will denote two particle state). The label b specifies the basis while the label $c$ gives the state within the basis. The above formula clearly accounts for viewing the states as lines in phase space: b specifies the orientation angle in phase space, while $c$ - its intercept, e.g. for $q=0$ the line intercept the p axis at $q={b \over a}$. The basis parametrized with ${1 \over {|b|}}\rightarrow 0$  is taken to be (see \cite{wootters3})  $\langle x|1;\pm\infty,c\rangle\;=\;\delta(x-c)$, these are eigenstates of the position operator and form our computational basis. The state-vectors,  for each basis parametrized by b, form a complete orthonormal basis \cite{wootters3}:
\begin{eqnarray}
                 \langle 1;b, c|1;b, c'\rangle\;&=&\;\delta(c-c'), \nonumber \\
\int {dc\left|1;b;c\rangle \langle 1;b,c \right|}\;&=& I.
\end{eqnarray}
 One may check directly that the vectors form an MUB set:
$$|\langle 1;b,c| 1;b',c'\rangle|\;\;independent\;of\;c,c'\;\;for\;b \neq b'.$$
The phase space geometry of MUB introduced by Wootters, \cite{wootters1}, may be seen
as follows. Two lines (cf. Eq.(\ref{wigner1})) parametrized with the same b but distinct c can never intersect. Two lines with distinct b's  intersect once.\\

Entanglement has been of great interest in theoretical physics since it first presentation by  Einstein Podolsky and Rosen  \cite{epr} where they consider an entangled two
particles' state  given by (the normalization is not specified)
\begin{equation}\label{epr}
|2 \rangle_{EPR}\;=\;\int {dx_1dx_2 \delta(x_1-x_2)|x_1\rangle |x_2 \rangle}.
\end{equation}

The ensuing studies led to the immense developement of quantum information theory  \cite{nielsen,diosi, peres2} and, concommitantly,  cryptography. On general grounds the sensitivity of quantum mechanical cryptographic protocols to eavsdroppers is anchored in a fundamental nature of quantum mechanics: a measuremant (e.g. by an eavesdropper) disturbs the system, i.e. leaves a trace (unless the state is an eigenstate of the operator representing the measurement). An essential ingredient is the use of non orthogonal states, e.g. vectors from distinct MUB. Indeed this was the basis of the pioneering protocol of encodement \cite{bb}.\\
A protocol wherein  cryptographic security is based on entanglement was put forward by Ekert, \cite{ekert}. Here, too, noncompatible (i.e. belonging to distict MUBs) bases are used. This idea involved qubits. It was extended with improved security to higher dimensional systems (qudits) \cite{peres1, cerf, durt}. The maximal number of MUB was shown \cite{tal} to be d+1. Ivanovic, \cite{ivanovich} demonstrated that for d=p (a prime) the set (i.e. d+1 bases) allows what is probably most efficient means for the determining the density matrix of an arbitrary state. This may be viewed as a facet in the study of d-dimensional quantum mechanics. It attracted a large body of research work with several cogent reviews \cite {vourdas1, vourdas2,
klimov1,saniga}. These studies now involve abstract algebra and projective geometry: \cite{wootters3,saniga,wootters4,bengtsson,klimov1,klimov2,planat1,planat2,combescure}.  A central issue is the unsolved problem of the number of MUB for  dimensionality, d, which is {\it not} a power of a prime. Of particular interest for the present work are the articles by Planat and coworkers \cite{planat1, planat2} who studied entangled states in conjuction with MUB sets similar to ours (see also \cite{klimov1, klimov2}).


In the present work we consider an interrelation between MUB and EPR like \cite{epr} entanglement. Our considerations are for continuous variables (i.e. $d \rightarrow \infty$) allowing us, thereby, to bypass some of the intricate problems that do arise in the finite dimensional case. In particualr we have Hermitian (i.e. measureable) operators that classify the MUB. We consider in the next section the one particle observable, $(\hat{p}-b\hat{x})$, whose measurement  projects on to Wootters' \cite{wootters1, wootters3} "line in phase space" state. The projection is elaborated in appendix A. A two-particle state is built via an EPR \cite{epr} like entanglement. (The corresponding finite dimensional case is considered in appendix B.) In the succeeding section we  give a secure protocol for (quantum) key distribution which, we believe, is conceptually simpler than available in the literature albeit is technically rather forbidding as it requires, for each transmitted bit two entangled states that are measured (prepared) at the encyphering port, A (Alice's), and then one (of each) of the entangled pair is brought to B (Bob) while retaining coherence with its mate. The last section includes conclusions and remarks while the structure of the entangled states and their relation to MUB is given in appendix C.

\section{Entangled Mutually Unbiased Bases}

The expression for the complete set of MUB for the continuous ($d \rightarrow \infty$)  case was given by Wootters \cite{wootters3}. We now recast it in our notations, the space location, x, is our "computational basis" (i.e. in terms of which we express all states):
\begin{equation}\label{xrep}
|1;b, c\rangle\;=\;{1 \over \sqrt {2\pi }}\int_{-\infty}^{\infty}dxe^{i[{b \over 2 }x^2+cx ]}|x\rangle .
\end{equation}

(The normalization of $\langle x|1;b, c\rangle$, given in \cite{wootters3}, was chosen to assure that $\int_{c_1}^{c_2}dc|1;b, c\rangle \langle 1;b, c|,\; c_2>c_1,$ is a projection operator \cite{wootters3}.)  EPR like entanglement, Eq.(\ref{epr}), leads to the following state  (a related entanglement was considered in \cite{planat1, planat2}),
\begin{equation}\label{entangled}
|2;b,c\rangle_{\alpha}={1 \over \sqrt {2\pi }}\int dx_1dx_2\delta(x_1-x_2)e^{i[{b \over 2 }x_1^2+cx_1 ]}|x_1\rangle |x_2\rangle_{\alpha} .
\end{equation}
$\alpha=1,2$ as we require two entangled states.  Thus we have for both states,
\begin{equation}\label{ent}
\langle x_1|\langle x_2|2;b,c\rangle={1 \over \sqrt {2\pi }}\delta (x_1-x_2) e^{i[{b \over 2 }x^2+cx ]}.
\end{equation}
This is the entangled state associated with the MUB labeled by b.  An essential ingredient of this expression is the additive attribute of the exponents, which in turn is a consequence of the states being of an MUB set (cf. Eq.(\ref{xrep}) and  the analysis in the appendix B):
$$e^{i[{b \over 2 }x^2+cx ]}=e^{i[{b_1 \over 2 }x^2+c_1x ]}e^{i[{b_2 \over 2 }x^2+c_2x ]}.$$

For $b=b_1+b_2,\;\;c_1+c_2=c.$  Now a measurement of the operator $(\hat{p}_1-b_1\hat{x}_1)$  with the state $|2;b,c\rangle$, Eq.(\ref{ent}), yielding the value $c_1$ in the A port, projects the state of the second particle,
 at the B port, to $|1;b_2, c_2\rangle $  (up to a trivial normalization constant). We outline the proof in appendix A. A finite dimensional version of this is given in appendix B.

\section{The Encryption Protocol}

We now give a protocol for a secure encryption. We consider an encrypted one bit message from A to B given that they share two entangled states ($\alpha\;=\;1,2$) cf. Eq.(\ref{ent}) both having the same values for the parameters b and c. In this case, of common values of b and c, we assume that neither party requires the actual values of the parameters. ( For the cases where the entangled states differ in their eigenvalues, c, Alice has to use the difference of these values to affect the encryption.)   The measurements relevant to this are the {\it two} particle observables ${\hat R}_1,{\hat R}_2$, Eq.(\ref{r}) discussed in Appendix C. The protocol is as follows: Alice measures for both states the quantity $(\hat{p}_1- b_1\hat{x}_1)$. She may choose any $b_1$ but use it for both states. She records the two values she gets: $c_1$ and $c'_1$.  She now sends, by classical communication channel, a value $\lambda$ to Bob's port. Upon receiving Alice's message Bob subjects his second particle to the unitary transformation $e^{i\lambda \hat {x}_2}$. This shifts his state from $c'_2=c-c'_1$ to $c'_2+\lambda$ (cf. Appendix A).  Now he measures the correlation among his two states,
\begin{equation}
|\langle 1;b_2,c_2|1;b_2,(c'_2+\lambda) \rangle|\;=\;\delta(c_2-c'_2-\lambda).
\end{equation}
Thus if Alice chooses to communicate +1 she sends $\lambda=c'_1-c_1,$ if she chooses to communicate zero she sends  $\lambda \neq c'_1-c_1$. This proceedure is secure in as much as Alice and Bob can at any time check, via classical communication channels, whether the observed value is indeed the intended one: An eavesdropper must use the same MUB measurement (i.e. the same $b_1$ chosen by A) and observe the same value $c_1$ in order not to leave a trace. In the case considered here both quantities range over unlimited numbers leaving him/her a small chance of avoiding detection.\\

\section{Conclusions and Remarks}
Mutual unbiased bases may be characterized by observables that are diagonal in each of them
(hence cannot be so in any two distinct bases). Einstein, Podolsky and Rosen (EPR) like entanglement of eigenfunctions of such observables has the property that measuring one particle of the entangled pair to be in a particular state (of the set of MUB bases' states) projects its (entangled) mate to another member state,  generally in another basis. This attribute allowed us to outline a secure (albeit technically complicated) protocol for encryption. The analysis led to the consideration of what might be termed mutually unbiased entangled pairs bases. The basis vectors in two particles systems are (maximally) entangled pairs whose labels are those of the MUB states used in the entanglement.  \\

\appendix{Appendix A: The Projection Proceedure}\\

We consider the Hermitian operator $(\hat{p}-b\hat{x})$. We denote its eigenfunctions by
$|1;b,c \rangle$:
\begin{equation}
(\hat{p}-b\hat{x})|1;b,c \rangle\;=\;c|1;b,c \rangle.
\end{equation}
We refer to the operator as the MUB operator (MUB-O). Its eignvalues for fixed b form a complete orthonormal basis where the various vectors in a given basis are labeled by c.  The definition of the MUB-O implies that \cite{wootters3},
\begin{equation}\label{mub1}
\langle x|1;b,c \rangle\;=\;\frac{1}{\sqrt{2\pi}} e^{i(\frac{b}{2}x^2+cx)}.
\end{equation}
The normalization is obtained in \cite{wootters3}. The MUB related
entanglement that we consider is given in Eq.(\ref{entangled}). Measuring at the A port an arbitrary MUB-O, e.g. with  a basis label $b_1$ and observing, say, $c_1,$ the projected state at the port B (before normalization) is:
\begin{eqnarray}\label{a3}
\langle 1;b_1,c_1 |2;b,c \rangle&=&\frac{1}{2\pi}\int dx e^{i(\frac{b_2}{2}x^2+c_{2} x)}|x\rangle \nonumber \\
 &=&\frac{1}{\sqrt{2\pi}}|1;b_2,c_2 \rangle,
\end{eqnarray}
with  $b_2=b-b_1;\;\;c_2=c-c_1.$ Thus the second particle is projected into a sharply defined, MUB labeled, state. \\

\appendix{Appendix B: Entaglement of MUB states at Finite dimensionality}\\

Our method may be applied in the finite dimensional cpace to the case where the states'
label are finite field variables. This is the case for dimensionality d, with $d=p^n,$ p an odd prime \cite{klimov1,klimov2,planat1,planat2}. The MUB are given by, using our notation, ($\omega_p=e^{i{2\pi \over p}}$):
\begin{equation} \label{mub}
 |1;b,c\rangle\;=\;{1 \over \sqrt{d}}\sum_{n\epsilon\mathbb{F}_d}{\omega_{p}^{tr[bn^2+cn]}}|n\rangle.
\end{equation}
Here $b,c,n\; \epsilon\; \mathbb{F}_d,$  $\mathbb{F}_d$ is Galois field of d dimension,
$|n\rangle$ denotes a vector in the computational basis (labelled with an element of the field); and $tr[\alpha]=\alpha+\alpha^p+\alpha^{p^2}+...+\alpha^{p^{n-1}}$. The trace ,tr, is a mapping with $\alpha \;\epsilon\;\mathbb{F}_d;\;\;tr{\alpha}\;\epsilon\;\mathbb{F}_p$.
Now a basic property of trace is: $tr[\alpha+\beta]=tr[\alpha]+tr[\beta]$ .Using our entanglement scheme \cite{planat1,planat2}, i.e. entangling within the computational basis, the entangled state corresponding to the generic MUB state, Eq. (\ref{mub}), is,
\begin{eqnarray}
|2;b,c\rangle&=&{1 \over \sqrt d}\sum_{n\epsilon\mathbb{F}_d}{\omega_p^{tr[bn^2+cn]}}|n\rangle
|n\rangle, \nonumber \\
 &=&{1 \over \sqrt d}\sum_{n\epsilon\mathbb{F}_d}{\omega_p^{tr[b_1n^2+c_1n]}
\omega_p^{tr[b_2n^2+c_2n]}}|n\rangle |n\rangle,
\end{eqnarray}

with $b_1+b_2=\;b,\;\;\;c_1+c_2=c.$
Thus projecting the first particle to any of the $b_1,\; c_1$ labeled state results in projecting the second particle to $|1;b_2,c_2 \rangle$ state in analogy with Eq.(\ref{a3}).
\begin{equation}
\langle 1;b_1,c_1||2;b,c\rangle\;=\;\frac{1}{\sqrt d}1;b_2,c_2\rangle.
\end{equation}

 No obvious measurement suggests itself for such projection. (Involvement of phase measurement by the Planat-Rosu phase operator  \cite{planat1} may be expected.)\\

\appendix{Appendix C: Structure of the Entangled States}\\

The determination, at the A (Alice's) port, of the two particle eigenvalue of the entangled state associated with the single particle MUB labeled by b - i.e. the detrmination of c of Eq. (\ref{entangled}) - requires operators whose measurement gives the eigenvalue while leaving the desired entangled state intact. The operators that characterize the (single particle) MUB are $\hat{p} - b\hat{x}$, Eq.(\ref{entangled}). Here b signifies the basis. To form an entangled state we require two {\it noncommuting} operators for {\it each} of the two particles. These are $\hat{p_1}-b_1\hat{x_1},\;\hat{p_1}-b_2\hat{x_1};\;\;b_1+b_2=b,\;\;b_1\neq b_2,$ for the first particle and  $\hat{p_2}-b'_1\hat{x_2},\;\hat{p_2}-b'_2\hat{x_2};\;\;b'_1+b'_2=b,\;\;b'_1\neq b'_2,$ for the second. These operators allow the definition of two {\it commuting} two particles' operators,
\begin{equation}\label{r}
\hat{R}_1=\hat{p}_1-b_1\hat{x}_1+\hat{p}_2-b_2\hat{x}_2;\;\;\hat{R}_2=\hat{p}_1-b'_1\hat{x}_1+\hat{p}_2-b'_2\hat{x}_2.
\end{equation}
The common eigenfunctions of these are, {\it necessarily} \cite{self}, entangled. The eigenfunctions of $\hat{R}_1,\;\hat{R}_2 $ are given by  Eq.(\ref{entangled}),
the eigenvalue is c. Thus the required operators are $\hat{R_1}, \hat{R_2}$.
To gain some insight to these states we  evaluate the Wigner function for the entangled state parametrized by b and c,
\begin{equation} \label{wigent}
W_{|2;b,c\rangle}(q_1,q_2,p_1,p_2)\;=\;\delta(q_2-q_1)\delta(p_1+p_2 -bq_1+c).
\end{equation}
Expressing thereby the relation of the entangled state to the MUB state bearing the same parameters.
 One of the authors(FCK)thanks NSERCC for financial suport.MR thanks the theoretical physics Institute for support during the stay at the University of Alberta.


\begin{thebibliography}{999}
\bibitem{self} We are unaware of any existing proof for this, rather obvious, observation. We may state the corresponding result for the finite dimension as follows. Let two operators on system $\alpha$ be $\mathbf{A},\; \mathbf{B},$ with $\mathbf{AB}=c\mathbf{BA},\;\;c\ne 1.$
Let two operators on $\beta$ be \textbf{U},\;\textbf{V}, with \textbf{UV}=d\textbf{VU}. Let the product of the c numbers cd=1. We have now that \textbf{AU} commute with \textbf{BV}. The statement is that the common eigenfunction of \textbf{AU} and \textbf{BV} is necessarily entangled.
\end{thebibliography}
\end{document}